# Managing the Public to Manage Data: Citizen Science and Astronomy

Peter Darch

Department of Information Studies,
University of California Los Angeles

## Abstract

Citizen Cyberscience Projects (CCPs) that recruit members of the public as volunteers to process and produce large datasets promise a great deal of benefits to scientists and science. However, if this promise is to be realised, and citizen science-produced datasets are to be widely used by scientists, it is essential that these datasets win the trust of the scientific community. This task of securing credibility involves, in part, applying standard scientific procedures to clean up datasets formed by volunteer contributions. However, the management of volunteers' behaviour in terms of how they contribute also plays a significant role in improving both the quality of individual contributions and the overall robustness of the resultant datasets. This can assist CCPs in securing a reputation for producing trustworthy datasets.

Through a case study of Galaxy Zoo, a CCP set up to generate datasets based on volunteer classifications of galaxy morphologies, this paper explores how those involved in running the project manage volunteers. In particular, it focuses on how methods for crediting volunteer contributions motivate volunteers to provide higher quality contributions and to behave in a way that better corresponds to statistical assumptions made when combining volunteer contributions into datasets. These methods have made a significant contribution to the success of the project in securing trust in these datasets, which have been well used by other scientists.

Implications for practice are then presented for CCPs, providing a list of considerations to guide choices regarding how to credit volunteer contributions to improve the quality and trustworthiness of citizen science-produced datasets.

# Introduction

Recent decades have seen an explosion of scientific sub-disciplines and projects which involve the processing of large quantities of data (Hey, Tansley, & Tolle, 2009; Welsh, Jirotka, & Gavaghan, 2006). Another critical challenge that has recently faced many scientists is a political culture which increasingly calls on scientists to show a willingness to engage the lay public with their work (Wynne & Felt, 2007).

To meet one or both of these significant challenges, some scientists have set up *Citizen Cyberscience Projects* (CCPs) (Grey, 2009). These CCPs are vast in terms of the scale of public participation. For example, the Zooniverse[1] (Lintott, 2010), a suite of 18 CCPs covering a range of scientific disciplines, has received contributions from nearly 900,000 volunteers. Another group of CCPs that all use the same piece of middleware called the *Berkeley Open Infrastructure for Networked Computing* (BOINC[2]) involves nearly a quarter of a million currently active volunteers.

One of the most critical challenges for scientists involved in running CCPs is securing credibility in the broader scientific community for citizen science-produced datasets (Riesch & Potter, 2013). To improve understandings of how a CCP's scientists might secure credibility for their project's scientific outputs, this paper presents findings from a case study of the first project in the Zooniverse. This is Galaxy Zoo, in the field of astronomy. The project was set up in 2007 with the aim of recruiting volunteers to classify the shapes of galaxies in approximately one million images taken from the Sloan Digital Sky Survey (SDSS) (Kent, 1994).

This paper will consider how the scientists and software engineers involved in the design, implementation and running of Galaxy Zoo have gone about the task of managing the processing of the dataset of SDSS images in order to produce scientifically credible datasets and discoveries of novel astronomical objects.

There is already a growing body of literature that discusses the data management practices of astronomers, including those involved in the storage and curation of SDSS data itself (Choudhury et al., 2007; Sands, Traweek, & Borgman, 2013) and astronomers who use SDSS data in their own work (Borgman, Sands, Wynholds, & Traweek, 2012b; Borgman, Wynholds, Wallis, Sands, & Traweek, 2012a). However, intrinsic to the data management challenges facing those who run Galaxy Zoo is another layer of management, namely that of the several hundred thousand members of the public who contribute to the project. It is upon this that the current paper focuses.

# Challenges for Citizen Cyberscience Projects

Projects that involve volunteer production of large datasets face a number of challenges. One is how to increase the quantity of the scientific work produced. This has already been widely considered elsewhere, both in terms of technical solutions for improving the efficiency of how volunteer contributions are used (Costa, Silva, Kelley, & Taylor, 2008; Kondo, Anderson, & McLeod, 2007) and social studies of volunteer motivations with a view improving volunteer retention (Darch & Carusi, 2010; Holohan & Garg, 2005; Krebs, 2010; Nov, Arazy, & Anderson, 2011; Raddick et al., 2010). Other challenges relate to ethical issues regarding how to credit volunteers' contributions, and how to keep volunteers updated on the scientific process (Riesch & Potter, 2013).

---

1  Zooniverse: https://www.zooniverse.org/
2  BOINC Stats: http://boincstats.com/en/stats/-1/project/detail/

A final challenge of great concern to scientists running CCPs is how to secure credibility for citizen science-produced datasets so that other scientists are willing to use these datasets (Riesch & Potter, 2013). Many large CCPs depend upon funding from public sources or foundations that support large scientific projects. In other words, these CCPs are in competition with other scientific projects for limited funds, and therefore must demonstrate that their output is of value to the scientific community.

### Trust and Credibility Issues with Datasets

Central to the decision of whether a particular scientist will use a dataset – citizen science-produced or otherwise – in their own work is trust in the reliability and quality of that dataset (van House, 2002). Firstly, it is critical that the scientist themselves trust the dataset. However, it is also very important that the dataset is trusted by the broader scientific discipline so that this scientist is able to receive scholarly recognition for conducting work that makes use of this dataset. Birnholtz and Bietz (2003) argue that datasets 'make a social contribution to the establishment and maintenance of communities of practice.' The reputation of a scientific discipline is bound up in part with the quality of datasets accepted as valid by that discipline: thus a dataset will not be widely regarded as trustworthy until scientists are convinced of its quality.

Assessing whether a dataset produced by another scientist is trustworthy is a complex process (Committee on Ensuring the Utility and Integrity of Research Data in a Digital Age, 2009). There are multiple factors that influence scientists' trust in datasets. One is the methods used to generate the dataset – are they scientifically reliable? Furthermore, is there sufficient documentation of these that allow for this reliability to be assessed (Faniel & Jacobsen, 2010)? Other factors relate to knowledge of who carried out the scientific work. Are they regarded as competent and producing high quality work (Zimmerman, 2008)? Are they regarded as honest (van House, 2002)?

Ensuring the reliability of a dataset involves the dataset's producers taking measures at every stage of the dataset's production (Mayernik, Wallis, Pepe, & Borgman, 2008). For these producers, securing others' trust in their datasets is thus a non-trivial accomplishment, even for datasets that have been generated by professional scientists (Wallis et al., 2007). However, in the case of citizen science-generated datasets, this has the potential to be an even more difficult task. First, the method of using lay members of the public to generate large scale datasets is something that scientists are generally not used to and could therefore be sceptical about. Second, the identities of volunteers are unknown, which complicates the issue of scientists determining whether those involved in creating the datasets are competent or honest. This is especially the case given that CCP volunteers cannot be assumed to have any scientific training.

### Securing the Credibility of Citizen Science-Produced Datasets

Some attention has been paid to improving the credibility of citizen science-produced datasets. Existing proposals are generally aimed at addressing the issue of 'problems of quality control and "bad data"' that can exist in citizen science (Edwards et al., 2013). In addition to technical solutions that seek to measure the reliability of volunteer-generated data (Fowler, Whyatt, Davies, & Ellis, 2013; Hunter, Alabri, & van Ingen, 2013) or to clean up volunteer-generated datasets (Fortson et al., 2012), Kamar and Horvitz (2012) recognise that improving dataset quality is also a task of managing volunteers. They propose a system of scoring volunteers based on how closely a volunteer's contribution matches the contributions of others. However, there can be a trade-off between speed and accuracy of volunteer contributions, and such a system would need to balance the two in a way that best meets the specific needs of a particular project. Thus, more work needs to be done to advance understandings of how to manage volunteers to improve the reliability of their contributions.

### Research Question

The above discussion motivates the following research question which will be addressed in this paper:

- What methods do scientists in a CCP employ to manage volunteers' behaviour in order to secure credibility for the volunteer-generated datasets?

# Case Study and Methods

This paper presents findings from a qualitative case study of Galaxy Zoo conducted between 2010 and 2012 as part of a doctoral dissertation (Darch, 2012). This section presents key features of Galaxy Zoo and outlines the methods used in the case study.

### Galaxy Zoo

Galaxy Zoo was launched with the aim of recruiting members of the public to classify nearly one million galaxy images produced by the Sloan Digital Sky Survey (SDSS), according to their morphological features. The first incarnation ran from July 2007, and the second phase was launched in February 2009. This second phase involved a more detailed classification scheme, using a subset of SDSS images. The Galaxy Zoo team then launched a third incarnation to classify images from the Hubble Space Telescope (HST), and a fourth was launched in October 2013 using images from a range of sources including SDSS, HST and the UK Infrared Telescope Deep Sky Survey[3].

The scientific work of Galaxy Zoo has two major components. The first is the *core work* of the project, which involves the classification of galaxy images and the production of statistical datasets generated by these classifications (for example, Lintott et al., 2011) which can then be used by astronomers external to the project to answer specific questions about galaxy formation and evolution. To date, the core work of Galaxy Zoo has resulted in at least 25 articles in peer reviewed astronomy journals.

The second component of Galaxy Zoo's scientific work comprises spin-off projects that have resulted from volunteers spotting unusual objects in galaxy images. These are known as *serendipitous discoveries*. One major example is that of *Hanny's Voorwerp* (Lintott et al., 2009), a novel astronomical object. This object became the subject of further investigation, and time on a number of telescopes was secured for observing the object. This has led to substantial coverage in the popular media.

---

3   The Story of Galaxy Zoo: http://www.galaxyzoo.org/#/story

## Case Study Methods

The case study followed the approach of 'virtual ethnography', which was developed to study communities mediated primarily or solely through the Internet (Hine, 2000). This involved conducting interviews with key people involved with the project, long term observations of the project's website (including the active online volunteer forums, and the project blog), and assembling and analysing a corpus of documents related to the project (such as publications, popular science articles, and proposals).

The interviews were semi-structured, and were conducted with 27 scientists, software engineers and volunteers involved with Galaxy Zoo. Interviews lasted between 45 and 150 minutes, with a median length of 60 minutes. They were recorded and transcribed. These transcripts, along with other documents, were analysed using grounded theory, which allowed for key themes to be identified (Glaser & Strauss, 1967). For further details on these methods, see Darch (2012).

# Findings

Securing the credibility of the Galaxy Zoo datasets has been a critical task for the scientists and software engineers involved in running the project (henceforth known as the *core team*). In particular, this is to ensure that other astronomers are interested in using these datasets in their own research because these other astronomers:

1. Need to enlist the support of funding bodies. To achieve this, these bodies need to be convinced of the credibility of the Galaxy Zoo data;

2. Are seeking to gain recognition of their peers, for instance through conference presentations or the publication of journal articles that are cited by other astronomers. This depends upon the wider credibility of Galaxy Zoo data.

### Securing Credibility with Other Scientists

The credibility of Galaxy Zoo data has been hard-won in the face of some scepticism in the broader scientific community, as recalled by a member of the core team:

> 'There were people who have spent years of their lives learning how to classify galaxies…so they're sceptical that some supposedly random person could just click a few buttons.'

The many painstaking steps taken to secure the credibility of the datasets bear testament to how important this has been for the project's core team. Measures used to clean up datasets are detailed in Fortson et al. (2012)[4]. These include: removing classifications from volunteers that were out of line with those of other volunteers (and thus suspected as malicious); introducing a weighting system for volunteers, with the classifications of volunteers who most closely match those of the other volunteers being assigned a higher weighting; multiple tests to assess whether volunteer classifications were systematically biased in any way, which has led to the development of methods to correct datasets for any such biases; and comparing the results of volunteer classifications with results generated by other methods in order to show that they agree with results generated by already credible methods.

---

[4] These methods of cleaning up datasets were implemented after the changes in managing volunteer behaviour discussed in this paper were made.

These methods all have something in common, namely that they have been applied to the datasets and classifications that have been provided by the volunteers: they do not involve the management of volunteer behaviour. However, the quality of Galaxy Zoo datasets also depends upon the behaviour of volunteers when classifying galaxies. A higher quality of individual classifications, as well as patterns of volunteer behaviour that better conform to statistical assumptions made in the analysis of these classifications, will in turn contribute to the production of more reliable datasets.

This case study has revealed that managing volunteer behaviour in a way that promotes a higher quality of Galaxy Zoo datasets has indeed been a central concern of Galaxy Zoo core team members. Furthermore, decisions and policies regarding how to publicly credit the contributions of volunteers have been central to the core team members' management of volunteer behaviour.

## Understanding Volunteer Motivations

Although the focus of this paper is on the decisions implemented by Galaxy Zoo core team members regarding volunteer management, this section will briefly discuss what motivates volunteers' participation in the project. This provides a background for understanding how core team members sought to manage volunteer behaviour. Raddick et al. (2010) analysed forum posts and conducted some interviews with Galaxy Zoo volunteers, finding that the three most frequently cited motivations for participation were an interest in astronomy, a desire to contribute to science, and a sense of awe at the vastness of the universe.

The first two motivations were strongly echoed in the interviews conducted for the case study reported here. However, those volunteers motivated by contributing to science often regard their potential contributions in different ways. For instance, some volunteers in particular value the fact that all volunteers are made to feel equally valuable to the project, and that all volunteers are making a contribution to the core scientific work, as exemplified by the following interview quotation:

> 'A zooite made a remark which I often quote: "The beauty of this site is that it is so level. There is no status attached to newbie, oldie, zooite, zookeeper[5]".'

Other volunteers reported being inspired by the possibility of making serendipitous discoveries, for example:

> 'Hanny's Voorwerp gives me hope that even someone like me can…make some sort of discovery.'

There were a range of other motivations that were expressed less frequently. These included friendships made through participating in the project online forums, the impact of the project on volunteers' offline lives (for example, one volunteer has been inspired by Galaxy Zoo to pursue a PhD), and enjoyment of seeing galaxy images.

---

5  "Zooite" and "zookeeper" are terms commonly used on the project forums to denote a volunteer and a core team member respectively.

## Methods of Crediting Volunteer Contributions in CCPs

Nov et al. (2011) divides Citizen Cyberscience Projects that involve the processing of large datasets into two broad categories, namely *high granularity* and *low granularity*. In low granularity projects, volunteers play a passive role in which chunks of data are downloaded onto their computer and processed using the computer's spare capacity. BOINC-based projects are an example of such projects. In high granularity projects, volunteers are asked to play a more active role by performing tasks on the project's website. Such tasks might involve categorizing galaxies from telescope images, or looking for particles in microscope images, or finding conformations of folded proteins. The Zooniverse projects are an example of high granularity projects.

Scientists running CCPs have a great deal of flexibility in terms of how they acknowledge volunteer contributions in public. In the case of low granularity projects, the primary method is usually to assign volunteers a score that relates to the amount of work conducted on their computers, and to produce tables that rank volunteers according to their score. This is common to all BOINC-based projects (Korpela, 2012).

In the case of high granularity projects, by contrast, there are a range of methods for acknowledging volunteer contributions. Some projects may assign scores to volunteers, and produce rankings. One example is FoldIt[6], where volunteers are presented with a protein and challenged to fold it into the lowest energy conformation that they can, and are scored according to the energy of the conformation they produce.

Galaxy Zoo, however, does not rank volunteers according to scores. When the Galaxy Zoo project was launched, the website did indeed include a league table ranking the top ten volunteers according to the number of galaxies classified. However, within two months of the project's launch, the decision was taken to abolish this table.

After this occurred, volunteer contributions to the Galaxy Zoo core work started to be acknowledged in other ways. Common to all the methods for crediting volunteer contributions that were subsequently implemented is that all volunteers are treated equally: either *all* volunteers are named in lists of contributors (such as on a page on the Galaxy Zoo website) or, in cases where a small subset of volunteers only are named (such as in conference oral or poster presentations), every volunteer has an equal chance of being mentioned, however many galaxies they may have classified.

A critical factor in moving away from a league table towards egalitarian forms of credit was to change volunteer behaviour in a way that would enhance the quality and credibility of the volunteer-produced datasets. In particular, there are three ways in which this is the case. These are considered in turn.

### Improving the quality of individual classifications

Galaxy Zoo scientists soon realised there was a trade-off between the speed and quality of classification, and that the existence of a league table was encouraging the former at the expense of the latter. Firstly, the league table communicated to volunteers that it was the volume, rather than quality, of classifications that was valued by the project. As a result, it was feared that the many volunteers motivated by contributing to science would take less care over each individual classification in order to classify more quickly. This is explained by a core team member:

> 'What we consciously don't want is a competition because we'd rather people classify well than classify as fast as they can.'

---

6   FoldIt: http://fold.it

It was also found that a league table was encouraging malicious behaviour: some volunteers were writing programmes that would send random classifications, and thus securing a high league table ranking but distorting classifications, as a core team member explained during an interview:

> 'I was surprised by how many people wanted to cheat the system, there were just kind of trolls sending bogus clicks, writing robots.'

As a result, abolishing the league table can be seen to influence volunteer behaviour to contribute to better quality datasets because volunteers are not encouraged to sacrifice quality for speed, and because it removes incentives for automated classifications.

**Improving the robustness of Galaxy Zoo data**

In addition to improving the quality of individual classifications, the core team also believed that a shift towards more egalitarian methods of crediting volunteers could assist in securing the credibility of their data in another way by bringing about:

1. A situation involving a very large number of volunteers each making a small number of classifications, rather than

2. A situation involving a small number of volunteers each classifying a large number of galaxies.

Situation 1 is believed to lead to greater robustness of the project's datasets than Situation 2. Two key assumptions of the statistical theory that has been used to process the data are that the number of classifications is very large and that each individual classification is independent of the others (and thus variations in the classifications are attributed to random statistical variation). Situation 1 (many independent volunteers making classifications) is much closer to meeting these assumptions than Situation 2, where the data might be skewed by systematic bias on the part of a single volunteer. The following quotation from an interview with a core team member reflects this:

> 'There are people who have classified one million galaxies, but that's not where our data comes from, data comes from people who have done 100… The reason why these classifications are so good is because we have 70 independent classifications of every galaxy.'

The shift from a league table towards egalitarian methods of crediting volunteers has been useful in the pursuit of Situation 1, in at least two ways. One way has been discussed above, namely that abolishing the league table would slow down the rate at which some volunteers were making classifications and thus reduce the potential skew if an individual volunteer exhibited systematic bias in their classifications.

The second was a fear that a league table was fostering the impression that their contributions were not important to achieving the project's scientific goals. As discussed above, a number of volunteers are motivated to continue participating in the project because they believe that the contributions of each volunteer – however small – are valuable to the project. Thus a league table was proving off putting to the broad mass of volunteers who contributed a relatively small number of classifications. For instance, a core team member stated in an interview:

     

'We didn't want people to feel that if they classify just a few galaxies then they're not really contributing.'

Thus, rather than singling out a very small group of volunteers for praise, as happened with the league table, it is important to make all volunteers feel as though they are making a genuine contribution to the project science, irrespective of how many galaxies they have classified. It can also be seen here why it was felt to be important not only to abolish the league table, but also to establish other methods of crediting volunteer contributions as positive statements that the contributions really are valuable.

### Potential for serendipitous discoveries

A third anticipated impact of having egalitarian forms of crediting volunteers discussed here is increasing the chance of volunteers making serendipitous discoveries by encouraging them to take more time over each classification. Although serendipitous discoveries can be understood as a scientific output of Galaxy Zoo in their own right, they can also be seen as helping to contribute indirectly to the credibility of the Galaxy Zoo datasets as well, in at least three ways.

One is that serendipitous discoveries have helped the project to locate itself within astronomy's mainstream. Serendipitous discoveries have played a critical role in the development of astronomy and thus the project has been able to continue this tradition, as explained by the following extract from an interview with a core team member:

'Some of the most interesting classes of objects [in astronomy] were initially discovered serendipitously. That's why I think serendipitous science is very important in astronomy.'

A second way in which serendipitous discoveries have helped to further the project's credibility is that the core team have been able to enrol other institutions and astronomers to study these discoveries, thereby endorsing the project's scientific value. A core team member recalls that:

'Outside interest in Hanny's Voorwerp was great for us, it really meant that other astronomers were starting to take the project seriously.'

Thirdly, the core team have become aware of the potential for serendipitous discoveries in the recruitment of more project volunteers. A larger set of volunteers is regarded by core team members as helping to promote a more statistically robust data set, and media coverage of serendipitous discoveries is usually accompanied by a spike in volunteer recruitment, as explained by a core team member during an interview:

'The media coverage has been really, really good. When there's a news story about the Zoo, we get an extra surge of and do more classifications.'

As discussed above, this was echoed in interviews with volunteers, some of whom stated they are indeed inspired to participate in the project by the prospect of making serendipitous discoveries.

**Credibility of Galaxy Zoo Datasets: Success**

A great deal of work has been put into securing the credibility of Galaxy Zoo datasets in the broader astronomical community. These efforts have largely been successful, as suggested by the following extract from an interview with a core team member:

> '[Scepticism] has decreased over the past year or two, and some people have got excited about it so people are coming and saying let's do a Galaxy Zoo with the cosmos or other surveys.'

In other words, not only has there been a broader acceptance of the Galaxy Zoo data, but the methods of Galaxy Zoo themselves have won credibility to the extent that other scientists are now seeing it as a viable way of generating data from other surveys.

The result of this is that funding bodies (such as the Leverhulme Foundation and the British government-funded Science and Technology Facilities Council) have been willing to support a number of proposals from teams of astronomers to work with the Galaxy Zoo data set. Such research has subsequently secured publication in peer reviewed journals and has, in turn, been cited by a wide range of cosmologists.

# Discussion and Implications for Practice

In Galaxy Zoo, a high granularity project, the abolition of the system of scoring and ranking volunteers according to the number of classifications made is associated with increasing the credibility of the Galaxy Zoo datasets, in three ways in particular. One is by slowing down the speed of volunteers' classifications to promote a higher quality of classifications, and reducing the incentive for malicious classifications, thus promoting honesty. The second is that it encourages a situation where classifications come from a very large number of volunteers, each making a relatively low number of classifications, thereby more closely matching statistical assumptions made when combining the volunteers' classifications into a dataset. Finally, by encouraging volunteers to take more time looking at each galaxy image, there have been serendipitous discoveries which have helped to locate the project in astronomy's mainstream, thereby enhancing the credibility of volunteer classification as a method for doing science.

Trusting in a dataset is an essential requirement for a scientist to use this dataset (van House, 2002). This trust is enhanced if the scientist believes that the methods used to produce the dataset are scientifically credible (Faniel & Jacobsen, 2010), if the individuals involved are regarded as producing high quality work (Zimmerman, 2008), and if these individuals can be regarded as honest (van House, 2002). It is thus perhaps unsurprising that Galaxy Zoo datasets are now largely trusted by astronomers.

**Implications for Practice**

Securing credibility for CCP-produced datasets is a non-trivial, yet highly important, task. Those who run CCPs need to make decisions regarding how to run their project in a way that promotes the credibility of the project's outputs. In addition to implementing technical solutions to cleaning up datasets generated by volunteers (such as those detailed in Fortson et al., 2012), they should be aware that the management of volunteer behaviour can make a significant contribution to these datasets' credibility. Methods for acknowledging volunteer contributions can have a major impact on this management, and there are a wide range of such methods available to those who run CCPs.

### Low granularity projects

As stated above, in low granularity projects, volunteers are commonly ranked according to how much work they have performed for the CCP. It is a straightforward task to calculate individual volunteers' scores based on the amount of data they have processed and the speed at which they process data has no impact on the quality of the outputs. Thus, in projects where there is no trade-off between quality and quantity of volunteer contributions, a league table can be seen as a way of increasing the scale of the scientific output without compromising quality (Darch & Carusi, 2010).

### High granularity projects

It might be seen from the example of Galaxy Zoo that, in the case of high granularity projects, volunteers' contributions should not receive some form of score. However, this is not so straightforward. It may be the case that it is possible to score the quality of a volunteer's contribution, such as in the case of the FoldIt project, where a better score corresponds directly to a better scientific contribution by the volunteer.

The case of FoldIt begs the question of whether a scoring system rewarding quality might improve the quality of volunteers' contributions in a high granularity project. Kamar and Horvitz (2012) proposed a scoring mechanism for CCPs that would reward volunteers depending on how closely their individual contribution matched the contributions of other volunteers. However, as has been seen the case of Galaxy Zoo, many members of the broader scientific community are initially sceptical of volunteer contributions as a method generating datasets. Thus, it is doubtful that using a citizen science-generated dataset as a benchmark against which to assess the quality of individual classifications can satisfy the sceptical broader scientific community.

### Issues to take into account when deciding how to credit volunteer contributions

The above discussion suggests that those involved in running a CCP and who are interested in securing credibility for their project-generated datasets should take into account the following issues when deciding how to credit volunteer contributions:

1. What is the nature of the task that volunteers are asked to do? Is there a trade-off between the quantity and quality of a volunteer's contributions?

2. Can the quality of the volunteer contribution be scored according to criteria that are well-established in the broader scientific community?

3. What assumptions are made when volunteer contributions are combined into a dataset? How can volunteer behaviour be managed to better correspond to these assumptions?

4. What other scientific outputs might there be from the project? Might these outputs enhance the scientific credibility of the project in general? And, if so, how might volunteer behaviour be managed to increase these outputs?

# Conclusion

It is vital for those who run a Citizen Cyberscience Project to consider how to secure credibility for the datasets being produced by their own CCP. This paper has considered how methods of crediting volunteers in the Galaxy Zoo project have played a significant role in volunteer management to improve reliability of Galaxy Zoo datasets. However, those running CCPs should also be aware that there are other features of CCPs, such as ways of reporting progress to the volunteers that can also impact upon volunteer behaviour (Darch, 2012). It is important to understand how these features can also guide volunteer contributions to better lead to the production of trustworthy datasets.

Furthermore, as scientific technologies develop, there are increasing data-related challenges for scientists. These provide changing opportunities and challenges for involving members of the public in the processing and production of datasets, and thus require further consideration of how best to manage volunteers. For example, future sky surveys, such as the Large Synoptic Survey Telescope (LSST), will produce data on a scale that vastly outstrips that of the Sloan Digital Sky Survey and will overwhelm the Galaxy Zoo project in its existing form (i.e. using volunteers to classify all of the images). However, in such circumstances, volunteer classifications might be used as training sets for automated image recognition methods in an iterative manner (Banerji et al., 2010).

Given current and likely future developments in the scope, scale and form of CCPs, it is therefore critical to improve understandings of how to manage volunteers to manage data if the promise of citizen cyberscience is to be fully-realised.

# Acknowledgements

I am deeply thankful to the Galaxy Zoo scientists, software engineers and volunteers for taking the time to speak with me. I would also like to thank Dr Annamaria Carusi (University of Copenhagen) and Dr Marina Jirotka (University of Oxford) for their support during my doctorate. Finally, I am grateful to the UCLA Knowledge Infrastructures project team for commenting on drafts of this paper: Professor Christine Borgman, Professor Sharon Traweek, Dr Jillian Wallis, Laura Wynholds, and Ashley Sands.